\RequirePackage{fix-cm}
\documentclass{svjour3}                     
\smartqed  
\usepackage{graphicx}
\usepackage{graphicx}
\usepackage{natbib}
\usepackage{color}
\usepackage{lineno}
\usepackage{float}
\usepackage[english]{babel}
\usepackage{appendix}
\usepackage{amsmath, amsfonts, amssymb, mathrsfs}
\begin{document}

\title{Chemical contamination mediated regime shifts in planktonic systems
}


\author{Swarnendu Banerjee         \and
        Bapi Saha \and
        Max Rietkerk \and
        Mara Baudena \and
        Joydev Chattopadhyay
}


\institute{S. Banerjee \at
              Agricultural and Ecological Research Unit, Indian Statistical Institute, 203, B.T. Road, Kolkata 700108, India \\
              \email{swarnendubanerjee92@gmail.com}           
           \and
           B. Saha \at
              Govt. College of Engineering and Textile Technology, Berhampore, West Bengal 742101, India
            \and
           M. Rietkerk \at
              Copernicus Institute of Sustainable Development, Utrecht University, PO Box 80115, 3508 TC Utrecht, the Netherlands
             \and
           M. Baudena \at
              Copernicus Institute of Sustainable Development, Utrecht University, PO Box 80115, 3508 TC Utrecht, the Netherlands\\\\
              Institute of Atmospheric Sciences and Climate, National Research Council (CNR- ISAC), Torino, Italy
              \and
           J. Chattopadhyay \at
               Agricultural and Ecological Research Unit, Indian Statistical Institute, 203, B.T. Road, Kolkata 700108, India
}

\date{Received: date / Accepted: date}

\maketitle

\begin{abstract}
Abrupt transitions leading to algal blooms are quite well known in aquatic ecosystems and have important implications for the environment. These ecosystem shifts have been largely attributed to nutrient dynamics and food web interactions. Contamination with heavy metals such as copper can modulate such ecological interactions which in turn may impact ecosystem functioning. Motivated by this, we explored the effect of copper enrichment on such regime shifts in planktonic systems. We integrated copper contamination to a minimal phytoplankton-zooplankton model which is known to demonstrate abrupt transitions between ecosystem states. Our results suggest that both the toxic and deficient concentration of copper in water bodies can lead to regime shift to an algal dominated alternative stable state. Further, interaction with fish density can also lead to collapse of population cycles thus leading to algal domination in the intermediate copper ranges. Environmental stochasticity may result in state transition much prior to the tipping point and there is a significant loss in the bimodality on increasing intensity and redness of noise. Finally, the impending state shifts due to contamination cannot be predicted by the generic early warning indicators unless the transition is close enough. Overall the study provides fresh impetus to explore regime shifts in ecosystems under the influence of anthropogenic changes like chemical contamination.

 \keywords{copper enrichment \and phytoplankton-zooplankton system \and alternative stable states \and stochasticity \and early warning signals}
\end{abstract}

\section{Introduction}
\label{intro}
Ecosystems can undergo abrupt transitions to an alternative stable state with fundamentally different characteristics \citep{scheffer2003catastrophic}. In lake ecosystems, such transitions, also known as regime shifts, are known to occur between turbid or algal dominated state and clear water state \citep{scheffer1993alternative}. Consequences of algal domination in water bodies include anoxic conditions leading to losses of fish and wildlife and also economic costs in the form of loss of recreational activities \citep{wilson1999economic,carpenter2008phosphorus}. These shifts are induced by interplay of several factors which includes food web interactions as well as nutrient dynamics \citep{scheffer1997ecology}. Over enrichment of water bodies with nutrients like phosphorus can enhance algal growth resulting in such blooms which are not easily reversible \citep{carpenter2005eutrophication}. However chemical pollutants like lipophilic substances and many metals can also negatively impact aquatic communities leading to species loss. Further, these pollutants may accumulate within the organisms through food and water and are passed through trophic interactions to higher levels in the food chains \citep{kooi2008sublethal}. The combined effect of how the pollutants internal concentration within these organisms might affect various life history traits and modulate ecological interactions determine the actual nature of ecosystem functioning \citep{huang2013model,kooi2008sublethal,garay2013more,huang2015impact}. Hence, a deeper insight into how contamination mediated alterations affect aquatic ecology is required to build a better understanding of regime shifts in contemporary world.

Industrial wastes and run-offs from agricultural fields often end up in water bodies thus making them prone to copper pollution \citep{jorgensen2010ecotoxicology}. Elevated copper concentrations in water bodies may have a toxic effect on several organisms \citep{flemming1989copper, clements1992assessment, WHOcopper} including both phytoplankton and zooplankton. Copper stress on phytoplankton cells negatively affect photosynthesis \citep{havens1994structural} and the concentration of chlorophyll \citep{fargavsova1999ecotoxicological}. Moreover, direct inhibition of growth has also been reported in many species due to bioaccumulation of the metal \citep{yan2002toxicity}. In model zooplankton species like \emph{Daphnia}, there are evidences that toxicity due to copper can lead to reduced body length \citep{knops2001alterations}, growth \citep{koivisto1992comparison} and survival \citep{ingersoll1982effect}. Nevertheless, copper is also biologically essential for most species \citep{mertz1981essential} and may lead to deficiency effects when present in very low concentrations \citep{bossuyt2003acclimation}. This hormetic dose-response relationship makes the study of copper even more interesting. Additionally, changing copper concentration can also modulate the trophic interaction of plankton in the food chain because of modification in behavioral traits like swimming velocity and mobility \citep{sullivan1983effects, gutierrez2012microcrustaceans}. 

Recent models of copper contamination have reinforced our understanding of plankton dynamics in polluted environments \citep{prosnier2015, camara2017copper, kim2018modeling}. Surprisingly, the impact on zooplankton predation by fish has been neglected in these studies, albeit fish density is a crucial factor in the context of water quality. Empirical studies \citep{mills1987fish, mcqueen1988cascading} indicate that zooplankton population collapse when fish density crosses a critical threshold (top down effect) \citep{luecke1990seasonal}. Trophic cascade via zooplankton allows the fish population to indirectly regulate the phytoplankton density. This has been captured by the classic minimal model by \cite{scheffer2000effects} which accounts for the complex nonlinearities involved in such interactions and the interplay between nutrient and fish density. The model demonstrates that a critical fish density can switch the ecosystem to the phytoplankton dominated state. Further predator-prey oscillations have been shown to favour the abrupt shift to phytoplankton domination. Although an earlier work by \cite{banerjee2019effect} took into consideration the effect of copper on fish predation, unfortunately it failed to study its impact on discontinuous transitions in planktonic systems.

Here, in this paper we will address this gap and also ask whether in the presence of predation pressure by fish, copper contamination can independently lead to such transitions in planktonic systems? Since, the importance of top down effects on planktonic regime shifts are known and both chemical influx and fish can be manipulated externally, with the help of a mathematical model, we attempt to understand the interplay between contamination and fish density. Environmental stochasticity often alters the system dynamics from that predicted by its deterministic counterpart \citep{dennis1998moving, hastings2004transients, baudena2007vegetation}. Especially, in the case of bistable models, stochasticity can induce or inhibit attractor switching the manner of which is not intuitive \citep{guttal2007impact, moller2009dynamic, sharma2015stochasticity}. Therefore we analyse the stochastic version of the our model and lastly investigate whether generic early warnings signals can predict the regime shifts in contaminated environment.

\section{Methods}
First, the original model by \cite{scheffer2000effects} is discussed briefly before incorporating the effect of copper enrichment in the system. A detailed description of the deterministic copper enriched model is provided followed by addition of stochasticity. Thereafter, the models are analyzed in order to understand how changing copper concentration influences ecological dynamics, especially regime shifts, in plankton ecosystem.


\subsection{Model description}
\label{sec:1}
The model by \cite{scheffer2000effects} is a two dimensional phytoplankton-zooplankton model as described below:

\begin{eqnarray} 
\centering
\small
\begin{array}{llll}
\displaystyle \frac{dP}{dt}  & = & \displaystyle rP(1-P/K)- \frac{aPZ}{k_P+P}+i(K-P), \\\\
\displaystyle \frac{dZ}{dt} & = & \displaystyle \chi \frac{aPZ}{k_P+P}- dZ-\frac{fZ^2}{k_Z^2+Z^2}\\
\label{plankton_model}
\end{array}
\end{eqnarray}

Here, $P$ and $Z$ denote phytoplankton and zooplankton densities respectively. The phytoplankton population is assumed to follow a logistic growth with carrying capacity $K$ and intrinsic growth rate $r$. They are predated upon by the zooplankton such that the predation follows a saturating functional response with the maximum predation rate denoted by $a$ and the half saturation constant denoted by $k_P$. Further, $\chi$ denotes the conversion efficiency of the predator and $d$ denotes its natural mortality rate. Since zooplankton forms only a part of the diet of many fish, the overall fish dynamics may not strongly depend on the zooplankton. Hence, it is reasonable to describe the fish predation as an additional mortality term in the zooplankton dynamics rather than considering explicitly the whole dynamics of fish population. This way the model complexity is also significantly reduced. The Holling type-III term used to model predation by fish can be interpreted as the average effect of many fish switching to foraging on zooplankton at different times. Here, $f$ denotes maximum predation by fish and $k_Z$ denotes the half saturation constant. The parameter $i$ in the last term of the first equation represents a diffusive inflow of phytoplankton. This is proportional to the difference between phytoplankton density in the part where the study is assumed to be carried out and the part where zooplankton are absent so that the phytoplankton are at carrying capacity. Although justification for the term has been discussed in detail in \cite{scheffer1995implications}, it is worth mentioning here that incorporation of such a stabilizing term brings the patterns generated by the minimal model close to biological reality \citep{scheffer2000effects}.  

In order to study how copper enrichment affects plankton dynamics, we adopt the approach by \cite{prosnier2015}. The effect of copper is introduced in the above system (\ref{plankton_model}) by multiplying each term in the model with the response of the associated trait to different copper concentrations. Such a response is denoted by $\varPsi_x$, where $x$ denotes the model parameter to which it is multiplied and the final model can be described as follows:

\begin{eqnarray} 
\centering
\small
\begin{array}{llll}
\displaystyle \frac{dP}{dt} & = & \displaystyle \varPsi_r \times rP(1-P/K)-\varPsi_a\times \frac{aPZ}{k_P+P}+i(K-P), \\\\
\displaystyle \frac{dZ}{dt} & = & \displaystyle \varPsi_a\times \chi \frac{aPZ}{k_P+P}-\varPsi_d\times dZ-\varPsi_f \times \frac{fZ^2}{k_Z^2+Z^2}\\
\label{final_model}
\end{array}
\end{eqnarray}

We make two important assumptions while including copper's effect on the model. First, the diffusion of phytoplankton in the system is not altered by the changing concentration of copper. Second, it is assumed that copper enrichment has neither any effect on phytoplankton carrying capacity nor on the half saturation constants of functional responses. It must be noted that $\varPsi_x$ are not parameters but they are dependent on copper concentrations. A particular trait of an organism would respond to the change in internal copper concentration, $C$. In the following paragraphs, we first explain how $\varPsi_x$ can be expressed as a function of $C$ for each parameter $x$. Further, if the copper concentration in the external environment is $E$, then the copper present within the organism will depend on the external environment and thus can be expressed as a function of $E$, i.e., $C(E)$. For analyzing our model, a specific functional form of $C(E)$ is required and so we describe in subsequent paragraphs how this function can be derived from a dynamic model of internal copper concentration.

\subsubsection{Modeling responses due to copper}
\label{sec:2}
\begin{figure}
\begin{center}
{\includegraphics[width= 1\textwidth]{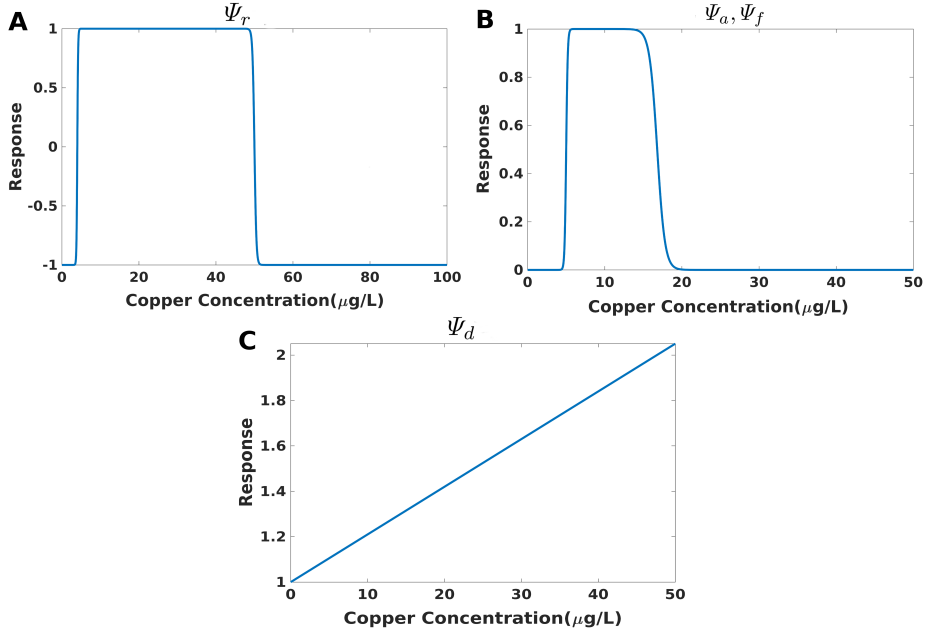}}
\end{center}
\caption{The effect of internal copper concentration $(C)$ on different parameters: (a) intrinsic growth rate ($\varPsi_r$), (b) maximum rate of predation ($\varPsi_a$), mortality due to fish predation ($\varPsi_f$), (c) natural mortality ($\varPsi_d$)}
\label{copper_response}
\end{figure}

From above, if the external copper concentration corresponding to higher $E_{C50}$ or toxicity is denoted by $h_x$ and lower $E_{C50}$ or deficiency is denoted by $l_x$, then the corresponding internal concentration is given by $C(h_x)$ and $C(l_x)$ respectively. Suffix $P$ and $Z$ have been used henceforth to identify the concerned organism as phytoplankton and zooplankton respectively. $m_x$ and $n_x$ are the positive and negative slope of the effect curve $\varPsi_x$. The effect of copper, which is vital for organisms but at the same time detrimental when present in large amount, can be captured using a asymmetric double sigmoid function (see Fig. \ref{copper_response}.A,B). For the algal intrinsic growth rate, the function  $(\varPsi_r)$, should range from $-1$ to $1$, where the maximum value is achieved for a range of intermediate copper concentrations. Such a curve must have negative values for high and low copper and can be expressed as follows \citep{prosnier2015} (Fig. \ref{copper_response}.A):\\

\begin{eqnarray}
\centering
\small
\begin{array}{llll}
\displaystyle \varPsi_r & = & \displaystyle -1 + tanh (m_r (C_P(E)-C_P(l_r)))- tanh (n_r (C_P(E)-C_P(h_r)))\\\\
\label{Cu_r}
\end{array}
\end{eqnarray}

There are hardly any empirical studies which connects copper concentration with grazing effort. However, it has been established that the movement of zooplankton like \emph{Daphnia} can be largely affected by copper concentration \citep{untersteiner2003behavioural, gutierrez2012microcrustaceans}. In particular, there is an increased movement at an intermediate copper concentration which is an advantage for predators like \emph{Daphnia}. As such, the effect curve of copper on grazing by zooplankton, $\varPsi_a$, can also be expressed as a double sigmoid function with maximum value 1 at the intermediate levels and minimum value 0 at toxic and deficient levels \citep{prosnier2015}(Fig. \ref{copper_response}.B):\\


\begin{eqnarray}
\centering
\small
\begin{array}{llll}
\displaystyle \varPsi_{a} & = & \displaystyle \frac{1}{2} tanh (m_a (C_Z(E)-C_Z(l_a)))- \frac{1}{2} tanh (n_a (C_Z(E)-C_Z(h_a)))\\\\
\end{array}
\end{eqnarray}

Although the increased mobility of the zooplankton helps them in their predation, it also increases their visibility to predators like fishes which mostly depend on visual cues to detect their prey \citep{wright1982differential, co1998swimming}. As such, the predation by fish is also maximum at intermediate copper and minimum when copper is toxic or deficient. Thus the effect curve, $\varPsi_f$ also ranges from 0 to 1 and it is assumed to be same as $\varPsi_a$ (Fig. \ref{copper_response}.B), i.e., $\varPsi_f$=$\varPsi_a$. Finally, since there is no evidence of decrease of mortality at high copper concentration, the effect on the death rate, $\varPsi_d$ can be modeled as a linear function which is as follows \citep{prosnier2015} (see Fig. \ref{copper_response}.C):

\begin{eqnarray}
\centering
\small
\begin{array}{llll}
\displaystyle \varPsi_d & = & \displaystyle 1+m_d\times C_Z(E)\\\\
\end{array}
\end{eqnarray}

\subsubsection{Modeling internal copper concentration}

In order to derive the function $C(E)$, we consider the rate of change of internal copper concentration for plankton which can be expressed with a model adapted from \cite{luoma2005metal}:

\begin{eqnarray}
\small
\centering
\begin{array}{llll}
\displaystyle \frac{dC(t)}{dt} & = & \displaystyle \frac{u_{m}E}{u_{c}+E}+ (C_{food}\times A_Z \times I_Z)-u_{e}\times C(t)
\end{array}
\label{biodynamic}
\end{eqnarray}

The first term on the right hand side represents bioaccumulation or direct uptake from environment. Such uptake is considered to be a saturating function of the copper present in the environment ($E$) where $u_m$ and $u_c$ are the maximal intake rate and half saturation constant respectively. This is plausible because of the competition that is present among the copper ions \citep{lebrun2012modelling}. The second term takes into account the intake from food which is responsible for biomagnification. Here, $C_{food}$ denotes the copper present in the food,  $I_Z$ denotes the ingestion rate of the zooplankton given by $\displaystyle \frac{aP}{k_P+P}$ and the assimilation efficiency is denoted by $A_Z$ which is equal to $\chi$ in our model. This term must not be present in the case of photosynthetic organisms like phytoplankton which does not prey upon another organism. The last term denotes the loss rate of the internal copper. Equating the right hand side of Equation (\ref{biodynamic}) to zero and in view of the above, one case easily calculate the steady state internal copper concentration for phytoplankton ($C_P$) and zooplankton ($C_Z$) in terms of $E$ \citep{prosnier2015}. An extra subscript $P$ or $Z$ is added to each of the parameter to indicate it's association with the particular organism.


\begin{eqnarray} 
\centering
\small
\begin{array}{llll}
\displaystyle C_P(E) & = & \displaystyle (\frac{E\times u_{mP}}{E+u_{cP}})\times \frac{1}{u_{eP}},\\\\
\displaystyle C_Z(E) & = & \displaystyle (\frac{E\times u_{mZ}}{E + u_{cZ}}+\chi\times \frac{a\times P}{k_P+P}\times C_P)\times \frac{1}{u_{eZ}}\\\\
\end{array}
\end{eqnarray}

\subsection{Stochastic model}

Random environmental fluctuations are crucial to the understanding of ecological system and may have consequences in its community stability and persistence. Moreover, the complexity induced by the
combined effect of stochasticity and nonlinearity in a system is fascinating and its investigation is especially warranted when there is bistability \citep{guttal2007impact}. As such, we explore the above system in the presence of stochasticity. Let the deterministic model (Equation \ref{final_model}) be denoted as:

\begin{eqnarray} 
\centering
\small
\begin{array}{llll}
\displaystyle \frac{dY}{dt}= G(Y)\\\\
\end{array}
\end{eqnarray}

Here, $Y=[P,Z]^T$ where $T$ denotes transpose of the vector and $G(Y)$ are the functions on the right hand side of the Equation \ref{final_model}. Here, we add an extrinsic multiplicative noise to the system after which it can be expressed as follows:

\begin{eqnarray} 
\centering
\small
\begin{array}{llll}
\displaystyle \frac{dY}{dt}= G(Y) + \sigma Y \xi(t)\\\\
\end{array}
\end{eqnarray}

where, $\sigma$ denotes intensity of noise and $\xi(t)$ denotes Gaussian white noise with zero mean and unit variance. While other alternatives like additive noise \citep{kefi2013early} can be used to model stochasticity, multiplying the noise term with the state variable is a more commonly used characterization of environmental fluctuations in ecological models \citep{evans2013stochastic, sharma2015stochasticity}. In this case, there is significantly reduced fluctuations around low density states and when the state is zero, the noise will vanish.

\begin{table}[h]
 \tiny
 \caption{Parameter values used in our simulation}
 \vspace{0.5 cm}
 \label{par_table}       
 \begin{tabular}{lllll}
 \hline\noalign{\smallskip}
 
Parameters & Value                  & Unit             & Description                                        & Reference\\
\noalign{\smallskip}\hline\noalign{\smallskip}
\multicolumn{2}{c}{Population dynamics}  \\
$K$                 &    3              &  $mgC L^{-1}$    &  algal carrying capacity                               &  \citep{murdoch1998plankton}  \\
$r$                 &    0.5            &  $d^{-1}$        &  algal intrinsic rate of natural increase              &       \\
$a$                 &    0.4            &  $d^{-1}$        &  maximum intake rate of \emph{Daphnia}                 &  The value of these              \\
$k_P$               &    0.6            &  $mgC L^{-1}$    &  half saturation constant of \emph{Daphnia}            &  parameters are same           \\
$\chi$              &    0.6            &    -             &  \emph{Daphnia} conversion efficiency                  &   as used in  \\
$k_Z$               &    0.5            &  $d^{-1}$        &  fish predation rate of susceptible \emph{Daphnia}     &    \cite{scheffer2000effects}                           \\
$f$                 &    0.1            &  $mgC L^{-1} d^{-1}$    &  fish predation rate                            &                               \\
$i$                 &    0.03           &  $d^{-1}$               & diffusive inflow of algae                       &    \\
$d$                 &    0.05           &  $d^{-1}$        &  \emph{Daphnia} natural mortality rate                 &   \citep{murdoch1998plankton}  \\
$\sigma$            &    0.035-0.045    &  $d^{-1}$        & noise intensity                                        &                                \\\\
\multicolumn{2}{c}{Copper concentration}  \\
$E$              &    0-100               &   $\mu g L^{-1}$          & range of external copper concentration      &  \\
$u_{mP}$          &    20                  &   $\mu g g^{-1} d^{-1}$   & algal maximal intake rate                   &  All parameters   \\
$u_{mZ}$          &    15                  &   $\mu g g^{-1} d^{-1}$   & \emph{Daphnia} maximal intake rate          &  related to copper internal \\
$u_{cP}$          &    6                   &   $\mu g L^{-1} $         & algal half saturation constant              &  concentration were   \\
$u_{cZ}$          &    7                   &   $\mu g L^{-1} $         & \emph{Daphnia} half saturation constant     &  taken from \\
$u_{eP}$          &    1                   &   $\mu g d^{-1}$          & constant loss rate for algae                &  \cite{prosnier2015} \\
$u_{eZ}$          &    1                   &   $\mu g d^{-1}$          & constant loss rate for \emph{Daphnia}       &    \\
                  &                        &                           &                                             &  \\
\multicolumn{2}{c}{Effect of copper}  \\
$l_r$             &    4                   &  $\mu g L^{-1}$    &  algal growth's deficiency $EC_{50}$  	     &    \\
$h_r$             &    50                  &  $\mu g L^{-1}$    &  algal growth's toxicity $EC_{50}$    	     &  All parameters related  \\
$m_r$             &    5                   &   -                &  copper effect on algal growth        	     &  to copper effects on   \\
$n_r$             &    2                   &   -                &  copper effect on algal growth        	     &  algae and \emph{Daphnia}  \\
$l_a$      &    5                   &   $\mu g L^{-1}$   &  \emph{Daphnia} predation's deficiency $EC_{50}$   &  predation/mortality  \\
$h_a$      &    16.8                &   $\mu g L^{-1}$   &  \emph{Daphnia} predation's toxicity $EC_{50}$     &   were taken from  \\
$m_a$      &    5                   &   -                &  copper effect on \emph{Daphnia} predation         &    \cite{prosnier2015}\\
$n_a$      &    1                   &   -                &  copper effect on \emph{Daphnia} predation         &    \\
$m_d$             &    0.021               &   $g \mu g^{-1}$   &  copper response coefficient for \emph{Daphnia} mortality  &    \\
\noalign{\smallskip}\hline
\end{tabular}
 \end{table}

\subsection{Analyses}
 
\subsubsection{Model parameterization and bifurcations}

While the parameters related to the population dynamics in plankton, were largely adapted from \cite{scheffer2000effects}, few parameters were also taken from an empirical study by \cite{murdoch1998plankton}. All parameters are chosen with respect to algae and the often studied zooplankton, \emph{Daphnia}. The values for model parameters related to modeling the internal copper concentration and effect of copper were obtained from an earlier study \citep{prosnier2015}. All model parameters with their values and descriptions are enlisted in Table \ref{par_table}. Analyses throughout this paper have been carried out using the same set of parameters as given therein unless stated otherwise.

We rely on bifurcation theory to study the asymptotic behavior of the model with respect to changes in parameters which imitates different environmental conditions. Abrupt changes in the dynamics of the system as a consequence of gradual shift of a parameter lead to a bifurcation. We examine the system for such bifurcations with respect to environmental copper concentration in the system $(E)$. Also, since the predation pressure by fish can be externally manipulated by altering harvesting strategies, it is important to understand the simultaneous impact of contamination and fishing on the planktonic system. For this, we carry out a two parameter bifurcation analysis with respect to $E$ and fish predation, $f$. The bifurcation diagrams were produced using numerical continuation software MATCONT \citep{dhooge2008new} in MATLAB environment.

\subsubsection{Stochastic simulations}

The stochastic model was also simulated in MATLAB using Euler Maruyama method \citep{higham2001algorithmic}. The time step of integration is $\bigtriangleup t$=1 where each unit time represents one day in our model. To analyze the stochastic model, we ran the simulation up to 20000 days and calculated the mean of last 5 years (1825 days). We used 10000 such realizations in order to plot the probability density of phytoplankton and zooplankton. Since, we are interested to understand the conditions at which the system shifts to phytoplankton dominated state, the initial condition for all stochastic simulations are zooplankton dominated $(P=0.5$ $mgCL^{-1}$, $Z=3$ $mgCL^{-1})$. Additionally we increase redness of noise in the system to understand how the dynamics of the system changes with increase in lag-1 autocorrelation. For this, we consider the Gaussian stochastic process $\xi$ to have a temporal autocorrelation following $1/f^{\beta}$ frequency spectrum. This has been suggested as a good model for many autocorrelated noise in biology including environmental fluctuations \citep{halley1996ecology}. To generate a stochastic signal with spectral exponent $\beta$, we use algorithm prescribed in \citep{stoyanov2011pink} where the extreme situation $\beta \rightarrow 0$ is a white noise and $\beta>0$ means red shifted or positively autocorrelated.

\subsubsection{Early warning signals}

Lastly, we also check the robustness of established early warning indicators in predicting critical transition in such a system. Early warning signals are statistical measures which precede some catastrophic transition. As the system approaches a bifurcation point, it is predicted that certain features of the time series like variance and autocorrelation increases. Although these signals were not originally developed to predict stochastic state shift, it has been recently debated upon whether the early warning indicators are relevant for stochasticity-induced attractor switching \citep{drake2013early, boettiger2013no}. So it is useful to investigate briefly the robustness of metric-based early warning indicators in this context. We used Early Warning Signal toolbox to analyze the simulated time series preceding a state shift \citep{dakos2012methods}. The time series were subjected to Gaussian detrending with bandwidth 25 before they were analyzed to calculate the autocorrelation at lag-1 and the standard deviation. The moving window chosen for calculating each of the metrics is half the size of the simulated time series.

\section{Results}

We begin by demonstrating the dynamics of plankton under changing copper enrichment. Since critical transition to the phytoplankton dominated state is known to be possible as a consequence of high fish density, we also investigate the dynamics due to interplay between changing copper concentration and fish density. Thereafter, the effect of stochasticity in the bistable region is discussed and the simulated time series is tested for early warning signals of regime shifts.

\subsection{Plankton dynamics under changing copper concentration}

\begin{figure}
\begin{center}
{\includegraphics[width= 0.95\textwidth]{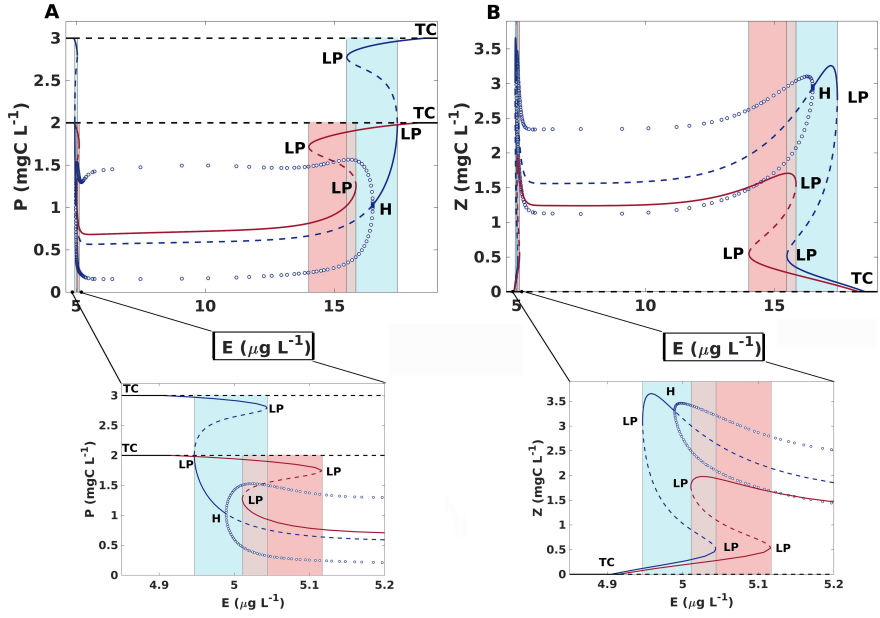}}
\end{center}
\caption{One parameter bifurcation diagram with respect to environmental copper concentration ($E$) for two different carrying capacities when $f=0.1$ $mgC L^{-1} d^{-1}$ . The interior equilibrium is denoted by the red lines for $K=2$ $mgCL^{-1}$ and blue lines for $K=3$ $mgCL^{-1}$. In both cases, the phytoplankton only equilibrium is denoted by black lines at $P=2$ $mgCL^{-1}$ and 3 $mgCL^{-1}$ respectively. The circles denotes the maximum and minimum amplitudes of oscillation while the solid and dashed lines denote stable and unstable equilibria. LP: Limit Point; TC: Transcritical; H: Hopf}
\label{K_2_K_3_bistability}
\end{figure}

The bifurcation diagram in Fig. \ref{K_2_K_3_bistability} demonstrates the change in densities of phytoplankton and zooplankton with respect to environmental copper concentration. Similar to the earlier studies \citep{prosnier2015, banerjee2019effect}, zooplankton ceases to exist via a transcritical bifurcation $(TC)$ when copper concentration is larger or smaller than a certain threshold. Additionally, we find that both toxic or deficient copper concentrations lead to a pair of limit points or fold bifurcations $(LP)$ resulting in bistability. Mathematically, such fold bifurcations are seen when the stable interior equilibrium collides with the unstable equilibrium as parameter passes through these points. These bifurcations lead to regime shifts whereby a small change in environmental copper concentration can result in an abrupt transition of ecosystem state from phytoplankton to zooplankton dominated state or vice-versa. Also, such changes are not easily reversible because the parameter must return much beyond the initial point of bifurcation for the system to return to its original state. 

We further investigate how different nutrient enrichment ($K$) might influence the manner in which copper affects the behavior of the system (see Fig. \ref{K_2_K_3_bistability}). For this, we track the bifurcations in the system with respect to environmental copper concentration for $K=2$ $mgCL^{-1}$ and $K=3$ $mgCL^{-1}$. In both cases, the phytoplankton density is lowest at intermediate copper concentration level and when moving towards toxic or deficient concentrations, there is bistability between phytoplankton dominated state and zooplankton dominated state. Additionally, when $K=3$ $mgCL^{-1}$, at intermediate concentration, the stable interior equilibrium loses its stability via Hopf bifurcation $(H)$ giving rise to population cycles. As a result of this, in certain parameter ranges the model also exhibits bistability between population cycles and phytoplankton dominance.

\begin{figure}[h]
\begin{center}
{\includegraphics[width= 1\textwidth]{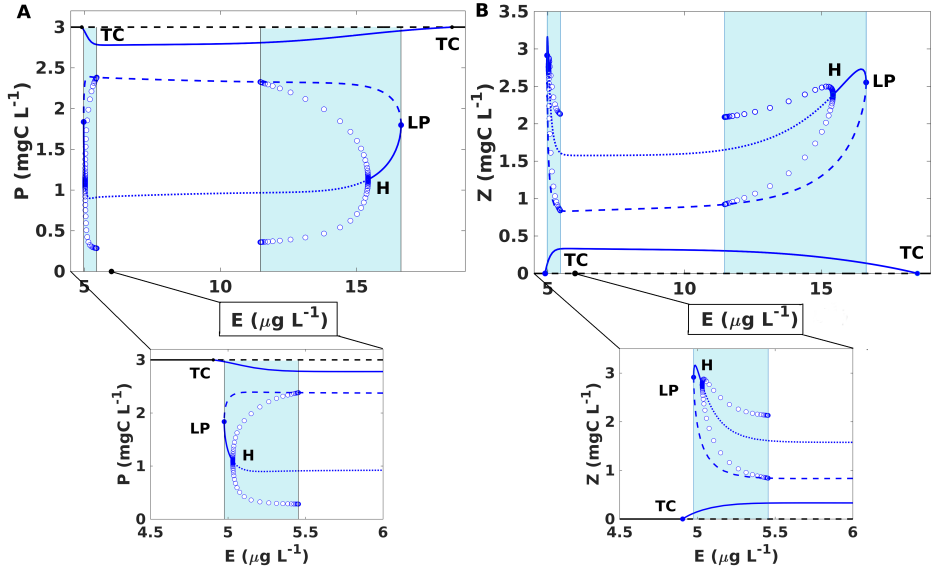}}
\end{center}
\caption{One parameter bifurcation diagram with respect to environmental copper concentration ($E$) when $f=0.15$ $mgC L^{-1} d^{-1}$ . The interior equilibrium is denoted by blue lines whereas the phytoplankton-only equilibrium is denoted by black lines. The maximum and minimum amplitudes of oscillation is denoted by circles while the solid and dash lines denote stable and unstable equilibria. LP: Limit Point; TC: Transcritical; H: Hopf}
\label{fish_hom}
\end{figure}

\subsection{Interplay between copper contamination and fish predation}
 
Since abrupt transition to phytoplankton dominated water has been attributed to zooplankton predation by fish beyond a critical threshold \citep{scheffer2000effects}, it is natural to ask whether the rate of fish predation can also have an impact on how the plankton dynamics might respond to changing environmental copper concentration. For this, we carry out the bifurcation analysis after increasing the fish predation rate to $f=0.15$ $mgC L^{-1} d^{-1}$ (see Fig. \ref{fish_hom}). From the extreme ends of the copper concentration axis, as we move towards the intermediate ranges, first there is bistability between population cycles and a phytoplankton-dominated equilibrium and then cycles grow in amplitude until it vanish abruptly via homoclinic bifurcations. Also, unlike the previous case in Fig. \ref{K_2_K_3_bistability}, a phytoplankton dominated state is always present and stable wherever a coexistence is possible thus leading to high phytoplankton density through out all ranges of copper concentration. 

\begin{figure}[H]
\begin{center}
{\includegraphics[width= 1\textwidth]{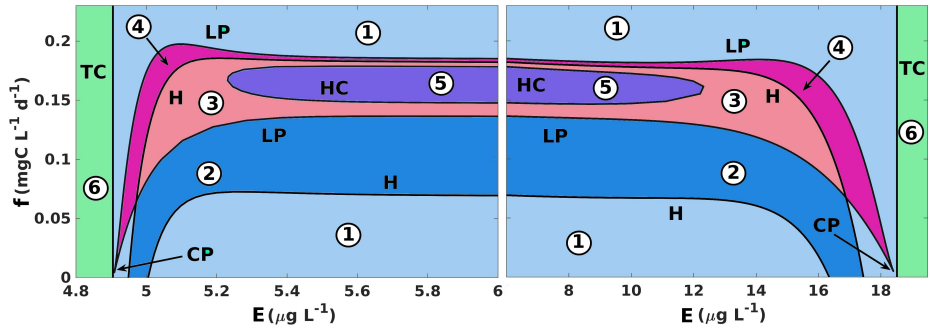}}
\end{center}
\caption{Two parameter bifurcation diagram with respect to copper concentration ($E$) and fish ($f$). Regions (1),(5): Stable coexistence, (2): Population cycle, (3): Bistability between stable coexistence and population cycle, (4): Bistability between two stable coexistence state, (6): Phytoplankton only state. All parameters are as given in Table \ref{par_table}.}
\label{two_parameter}
\end{figure}

In view of the above unintuitive results, it appears necessary to understand the complete dynamics exhibited by the system at different external copper concentration and fish predation pressure. The interplay between the two is demonstrated in Fig. \ref{two_parameter}. The system is in stable coexistence equilibrium in region \textcircled{1} when the predation rate, $f$, is very high. On reducing the predation by fish, which is equivalent to reducing fish density, the system becomes unstable leading to oscillatory dynamics. The population cycles can be observed in region \textcircled{2} and \textcircled{3} bounded on both sides by Hopf bifurcation lines $(H)$. In region \textcircled{3}, along with the oscillatory dynamics, depending on initial conditions, the system may also converge to phytoplankton dominated state. Embedded within \textcircled{3}, is region \textcircled{5} where the system has only one stable state which is phytoplankton dominated in nature. This bistable system dynamics extends to region \textcircled{4}, where the system can switch between two stable coexistence equilibria: high phytoplankton - low zooplankton density and low phytoplankton - high zooplankton density state. Mathematically, such a behaviour arise due to two cusp points $(CP)$ which are observed in both toxic and deficient copper regimes. When copper concentration is too low or high, the system undergoes transcritical bifurcation so that the zooplankton population becomes extinct and only the phytoplankton is able to survive in region \textcircled{6}.
 
\subsection{Effect of stochasticity in the bistable regime}

\begin{figure}[H]
\begin{center}
{\includegraphics[width= 1\textwidth]{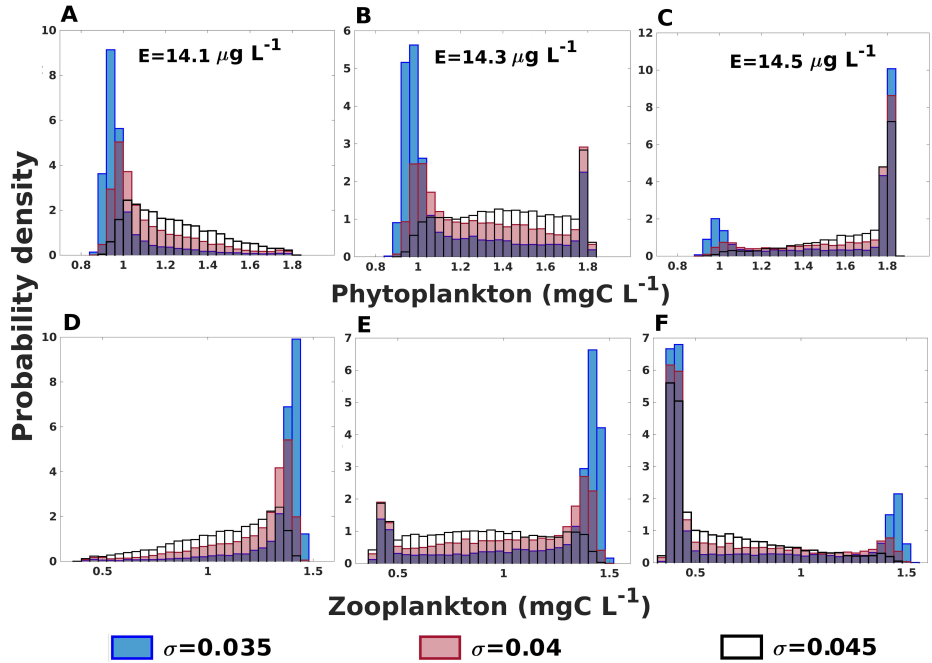}}
\end{center}
\caption{Probability density estimates of the phytoplankton and zooplankton populations under three different external copper concentration from 14.1 $\mu g L^{-1}$, 14.3 $\mu g L^{-1}$ and 14.5 $\mu g L^{-1}$; K=2 $mgCL^{-1}$. }
\label{K_2_stochastic}
\end{figure}

Although deterministic models are easy to analyze, in reality however, ecological systems are subject to environmental fluctuations and uncertainty that may be captured by a stochastic noise term. It is particularly interesting to study how environmental fluctuations influence a system with alternative stable states. When carrying capacity, $K$, is set to 2 $mgCL^{-1}$ we investigate the probability with which different values of phytoplankton and zooplankton densities are observed under three different external copper concentrations (see Fig. \ref{K_2_stochastic}). The concentrations were chosen such that they lead to bistability between phytoplankton and zooplankton dominated state in the deterministic set up and noise with intensity $\sigma=0.04$ $d^{-1}$ was used. When the environmental copper concentration, $E$, was 14.1 $\mu g L^{-1}$ and $E=14.5$ $\mu g L^{-1}$, the probability density was unimodal with the mode around the zooplankton dominated and phytoplankton dominated state respectively. Here, the shift to phytoplankton domination occurs much prior compared to the deterministic model where the system tips only at $E=15.84$ $\mu g L^{-1}$. An increased noise intensity ($\sigma=0.045$ $d^{-1}$) under such conditions lead to a decreased skewness of the probability density and vice versa. In between these two scenarios, when copper concentration is 14.3 $\mu g L^{-1}$, the density becomes bimodal where the system has almost equal chance to end up in phytoplankton dominated state or zooplankton dominated state. This bimodality is however lost also at this intermediate $E$ value when the noise intensity is increased. In the remaining sections we only focus on the effect of stochasticity in the toxic copper concentration ranges but the same analyses can also be carried out for the bistability range of low copper as well. In fact, a similar behaviour as described above, was demonstrated on addition of stochasticity in the deficient copper concentrations (see Appendix \ref{appA}). 

\begin{figure}[H]
\begin{center}
{\includegraphics[width= 1\textwidth]{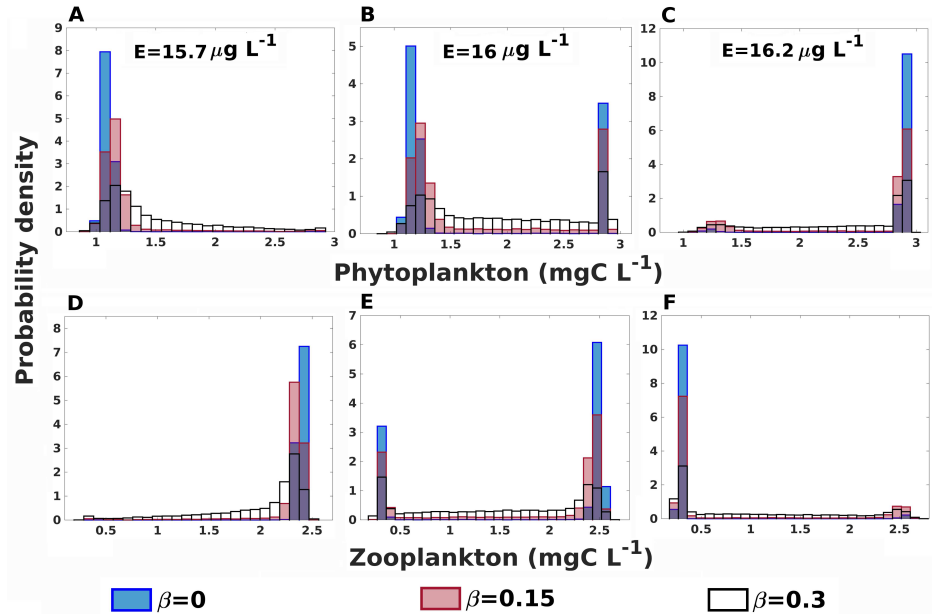}}
\end{center} 
\caption{Probability density estimates of the phytoplankton and zooplankton populations under three different external copper concentration from 15.7 $\mu g L^{-1}$, 16 $\mu g L^{-1}$ and 16.2 $\mu g L^{-1}$; K=3 $mgCL^{-1}$}
\label{K_3_stochastic}
\end{figure}

Since the phytoplankton dominated state in the previous analyses is close to the environmental carrying capacity of the algae, it is only natural to ask what impact does stochasticity have when the environmental carrying capacity is increased due to nutrient enrichment. In order to answer this, we investigated the effect of noise with intensity $\sigma=0.04$ $d^{-1}$ in the bistable regime when $K=3$ $mgCL^{-1}$ (Fig. \ref{K_3_stochastic}). We observe a similar transition from unimodal peak around the zooplankton dominated equilibrium at copper concentration 15.7 $\mu g L^{-1}$ to phytoplankton dominated state at copper concentration 16.2 $\mu g L^{-1}$. At the intermediate concentration, $E$=16 $\mu g L^{-1}$, we observe bimodal peak where the system has equal probability to converge to either low or high phytoplankton density. Thereafter, we increase the redness of the noise to $\beta=0.15$ and then $\beta=0.3$ and subsequently compare the probability densities to the $\beta=0$ case for all the three copper concentrations (Fig. \ref{K_3_stochastic}). Our results show that, similar to that of increasing noise intensity, red shifted noise also decrease the skewness of the probability densities. Moreover, in case where there was a bimodal density in presence of white noise, increasing redness leads to significant reduction in the peak height. 

For the bimodal cases corresponding to both the carrying capacities (Fig. \ref{K_2_stochastic}, B,E and Fig. \ref{K_3_stochastic}, B,E), looking at a specific simulated time series of the stochastic model, we observe stochastic switching between the equilibria (Fig. \ref{switch_stochastic}). In the case when $K=2$ $mgCL^{-1}$, multiple switch is observed unlike that of the other case where once the system switches to a phytoplankton dominated equilibrium, it is unable to return back. 

\subsection{Robustness of early warning signals}
\begin{figure}[H]
\begin{center}
{\includegraphics[width= 1\textwidth]{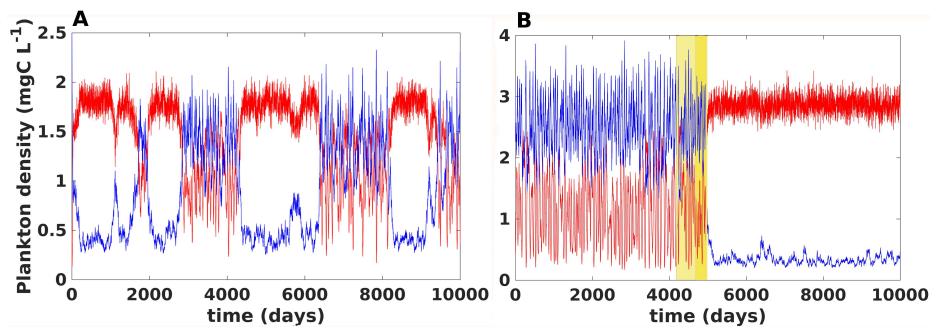}}
\end{center}
\caption{Time series simulation of the stochastic model for (A) $K=2$ $mgCL^{-1}$, $E=14.3$ $\mu g L^{-1}$ and (B) $K=3$ $mgCL^{-1}$, $E=16$ $\mu g L^{-1}$. The red and the blue line denotes the phytoplankton and the zooplankton densities respectively. Yellow shades represent the time segments that have been analyzed for early warning signals in Fig. \ref{ews}. }
\label{switch_stochastic}
\end{figure}

\begin{figure}[h]
\begin{center}
{\includegraphics[width= 1\textwidth]{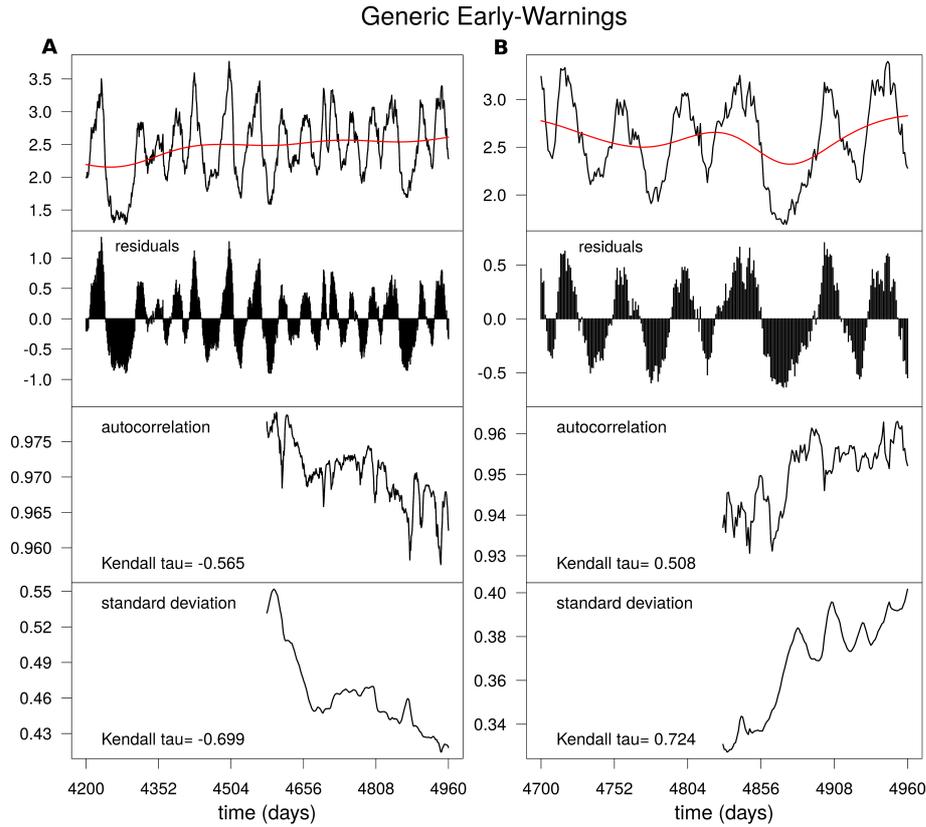}}
\end{center}
\caption{Early warning signals for a simulation of the stochastic model. Two time segments of different lengths (A) 4200-4960 days and (B) 4700-4960 days which precedes the regime shift from zooplankton dominated state to phytoplankton dominated state were analyzed.}
\label{ews}
\end{figure}

Since the switch to the phytoplankton dominated state is most likely irreversible when $K=3$, we analyzed the portion of the time series prior to such a shift denoted by the yellow shaded region in Fig. \ref{switch_stochastic}.B. Zooplankton biomass is known to provide early warning of regime shifts in lake community composition \citep{pace2013zooplankton}. As such the simulated data of zooplankton density was considered for a stretch starting from 4200 days to 4960 days (Fig. \ref{ews}.A). Both the metric, autocorrelation lag-1 as well as standard deviation decreases with time thus being unable to provide any early warning to the regime shift which we know occurred immediately after this time segment. Next, we analyze a shorter time series segment from 4700 to 4960 days (Fig. \ref{ews}.B) denoted by darker yellow shade in Fig. \ref{switch_stochastic}.B. Evidently, our analysis shows that the early warning signals performed much better in case of short time series segments.

\section{Discussion}

Abrupt transitions to phytoplankton dominated turbid water are known to occur in lake ecosystems but the impact of chemical pollution on such state shifts is not well understood. Fish density has been known to be an important driver of regime shifts in plankton community. To this end, using the approach prescribed in \cite{prosnier2015}, we introduced here a new model which takes into account both sigmoidal functional response for zooplankton predation by fish and its alteration under variable copper concentrations. Our analyses leads to a comprehensive understanding of the ecological dynamics due to the interaction between copper contamination and fish density. Further, consideration of environmental fluctuations in the form of stochasticity led to a clearer insight into how changing copper concentration of lake water may influence sudden shift to phytoplankton domination. 

Moving towards the extreme ends of the copper concentration axis, first the zooplankton ceases to exist followed by the phytoplankton. This is attributed to decreased consumption by zooplankton and the inhibition of phytoplankton growth in these ranges. Although this behaviour exhibited by our model is consistent with the earlier works \citep{prosnier2015, banerjee2019effect}, there is a notable change in plankton dynamics when both the functional groups coexist. Within such ranges, the earlier study by \cite{banerjee2019effect} demonstrated that toxic or deficient copper concentration could lead to destabilization of the predator-prey dynamics. However, on incorporating the sigmoidal functional response for fish predation, we find no such destabilizing behavior in these ranges. Instead a bistable scenario is observed where the system can switch to a phytoplankton dominated state (Fig. \ref{K_2_K_3_bistability}). This is counterintuitive as fish predation decreases in these ranges and so zooplankton dominance is expected. However, it must be noted that consumption by zooplankton also decreases in toxic or deficient concentrations thus leading to decline in its density and transition to phytoplankton dominated state. In the intermediate regimes, the stable coexistence equilibrium is characterized by low phytoplankton and high zooplankton density because of increased zooplankton predation. When carrying capacity is increased here, the system may retain a relatively low phytoplankton density but only via population cycles (Fig. \ref{K_2_K_3_bistability}). Such destabilization of the system occurs due to increase in energy flux from the phytoplankton to the zooplankton relative to the zooplankton's loss rate, as indicated by the ecological theory on stability \citep{rip2011cross}. Although oscillatory dynamics in intermediate ranges have been observed in earlier studies of copper enrichment \citep{prosnier2015, banerjee2019effect}, here it resulted in bistability between population cycles and phytoplankton dominated steady state which was not reported earlier. 

The complexity of the dynamics exhibited by the system can be better understood by studying the interaction between fish density and copper enrichment (Fig. \ref{two_parameter}). The above mentioned bistability which was observed at specific levels of copper enrichment, vanishes when fish density is comparatively high. Increase in zooplankton mortality at higher fish predation leads to stabilization of population cycles and phytoplankton domination across all copper concentrations. At intermediate fish density, the oscillatory dynamics may lead to collapse of the zooplankton population due to food shortage \citep{scheffer2000effects}. Mathematically this occurs via homoclinic bifurcation resulting in phytoplankton domination being the only stable state in the middle ranges of copper concentration. This is significant from an ecological point of view because under such parameteric ranges, once the system reaches the condition of phytoplankton domination, changing copper concentrations has no effect on the ecosystem state. However, maintaining copper concentration at proper levels, the chance of switch to a phytoplankton dominated equilibrium can be reduced (Fig. \ref{fish_hom}). When fish density is very low, the oscillations ceases and coexistence equilibrium is stable for a large range of copper concentration. This can be attributed to the diffusive inflow term which is stabilizing in nature (Fig. \ref{two_parameter}).

When stochasticity is added to the system, the bistability is weakened as the dynamics spends most of the time near the phytoplankton dominated state. In fact, starting from a zooplankton dominated condition, the system may become phytoplankton dominated much prior to the fold bifurcation. Only a very small range of copper concentration parameter demonstrates bimodality in the probability density of the observed values. This bimodality is lost on increasing noise intensity or redness. It is interesting to note here that, for a higher carrying capacity value, the bimodality is more prominent as the system can only be at the two extremes which in the deterministic set up represent a population cycle near lower phytoplankton equilibrium and a phytoplankton dominated equilibrium (see Fig. \ref{K_2_K_3_bistability}). At any point of time after a long run, the probability that the system displays intermediate values is very low. This happens mainly because for high carrying capacity, the boundary separating the basin of attractions is sufficiently distant from the two equilibria. As a result, once the system has switched to an alternate equilibrium, it is unable to switch back and continue in the same state unless the system is perturbed by noise of sufficiently large strength. In contrast, when the carrying capacity is comparatively low, both the phytoplankton dominated and zooplankton dominated equilibria are close to borderline separating the two basin of attractions thus allowing a frequent switch back and forth (Fig. \ref{switch_stochastic} A, B and Appendix \ref{appB}, Fig. \ref{boa},).

Analyzing the time series prior to such regime shift, we find that generic measures like autocorrelation and variance failed to indicate the approaching state shift. This is not remarkably unexpected provided the fact that it has already been argued that such measures were primarily developed to predict bifurcations and not stochasticity induced state shift \citep{boettiger2013no}. Nevertheless, the early warning indicators performed relatively well where a shorter time segment was analyzed. Similar results were reported in other bistable ecological systems which demonstrate noise induced regime shift \citep{sharma2015stochasticity}. However, it has been argued that it is not appropriate to conclude that the signals were successful because they failed to predict the upcoming transitions longer ahead. This failure to predict impending shifts further highlights the importance of the present study.

Summing up, our work reveals how both copper pollution as well as deficiency of copper can bring about regime shifts in lake ecosystems. Thereby, we want to stress the importance of deeper understanding of how human driven factors like nutrient enrichment, fishing and chemical pollution can interact with the complex ecosystem dynamics to bring about undesirable outcomes. The study also points to the importance of considering stochasticity in such modeling efforts as noise can influence the outcome which might be quite different from what is predicted from the deterministic model. Our results suggest that more efforts to understand the nonlinearity involving complex anthropogenic changes and their interaction with stochasticity is required to get a better insight into the present day scenario.

\section*{Acknowledgement}
Swarnendu Banerjee acknowledges Senior Research Fellowship from Council of Scientific and Industrial Research, India. The authors would also like to thank Hil Meijer, University of Twente for confirming the MATCONT simulations for the two parameter bifurcation diagram.

\newpage


%
%


\appendix

\section{Effect of stochasticity in low copper concentrations}
\label{appA}

\begin{figure}[H]
\begin{center}
{\includegraphics[width= 1\textwidth]{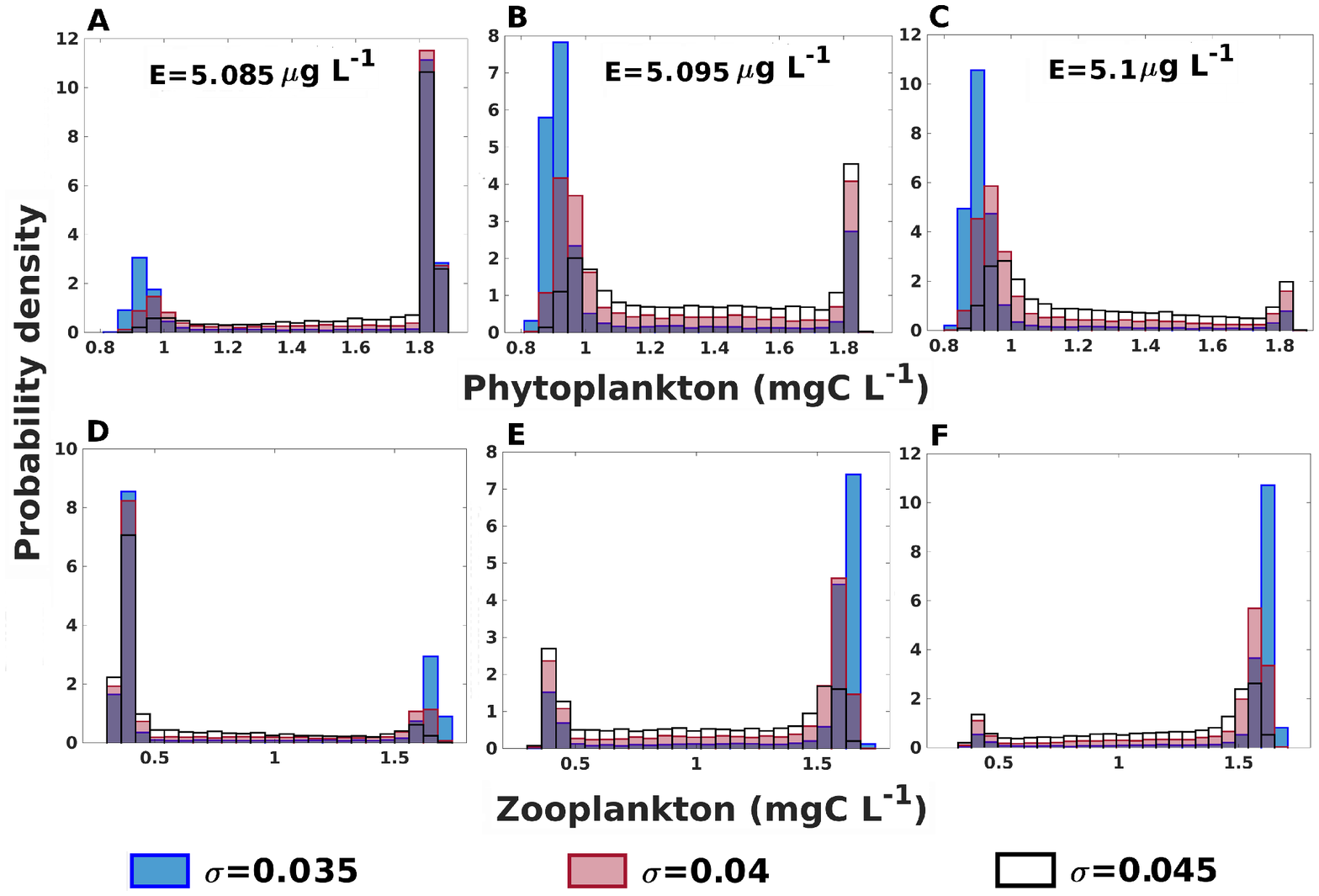}}
\end{center}
\caption{Probability density estimates of the phytoplankton and zooplankton populations under three different external copper concentration from 5.085 $\mu g L^{-1}$, 5.095 $\mu g L^{-1}$ and 5.1 $\mu g L^{-1}$; $K=2$ $mgCL^{-1}$. }
\label{K_2_stochastic_low}
\end{figure}

Deficient copper concentrations also lead to bistable system dynamics resulting in planktonic regime shifts. The effect of stochasticity on such low ranges of copper concentration is examined when carrying capacity $K=2$. Similar to the toxic concentration case, the system switches to phytoplankton dominated state prior to the fold bifurcation. The probability density of the observed values from the simulation is unimodal with mode around zooplankton dominated equilibrium at copper concentration 5.1 $\mu g L^{-1}$. Subsequent small decrease of copper results in the system demonstrating bimodality at concentration 5.095 $\mu g L^{-1}$ and unimodal mode around phytoplankton dominated state at concentration 5.085 $\mu g L^{-1}$ (see Fig. \ref{K_2_stochastic}). Increased intensity of noise leads to decreased skewness of the probability densities.

\section{Basin of attraction for the alternative stable states}
\label{appB}

The stochastic switch between the attractors in Fig. \ref{switch_stochastic} can be understood with the help of basin of attraction for the two equilibria under different carrying capacities. When $K=2$, the boundary separating the basin of attraction is very close to both the phytoplankton and zooplankton dominated equilibrium which facilitates multiple stochastic switching. On the other hand, the boundary is relatively farther away from the two attractor in case of higher carrying capacity, i.e., $K=3$ resulting in very infrequent switch. 

\begin{figure}[H]
\begin{center}
{\includegraphics[width= 0.8\textwidth]{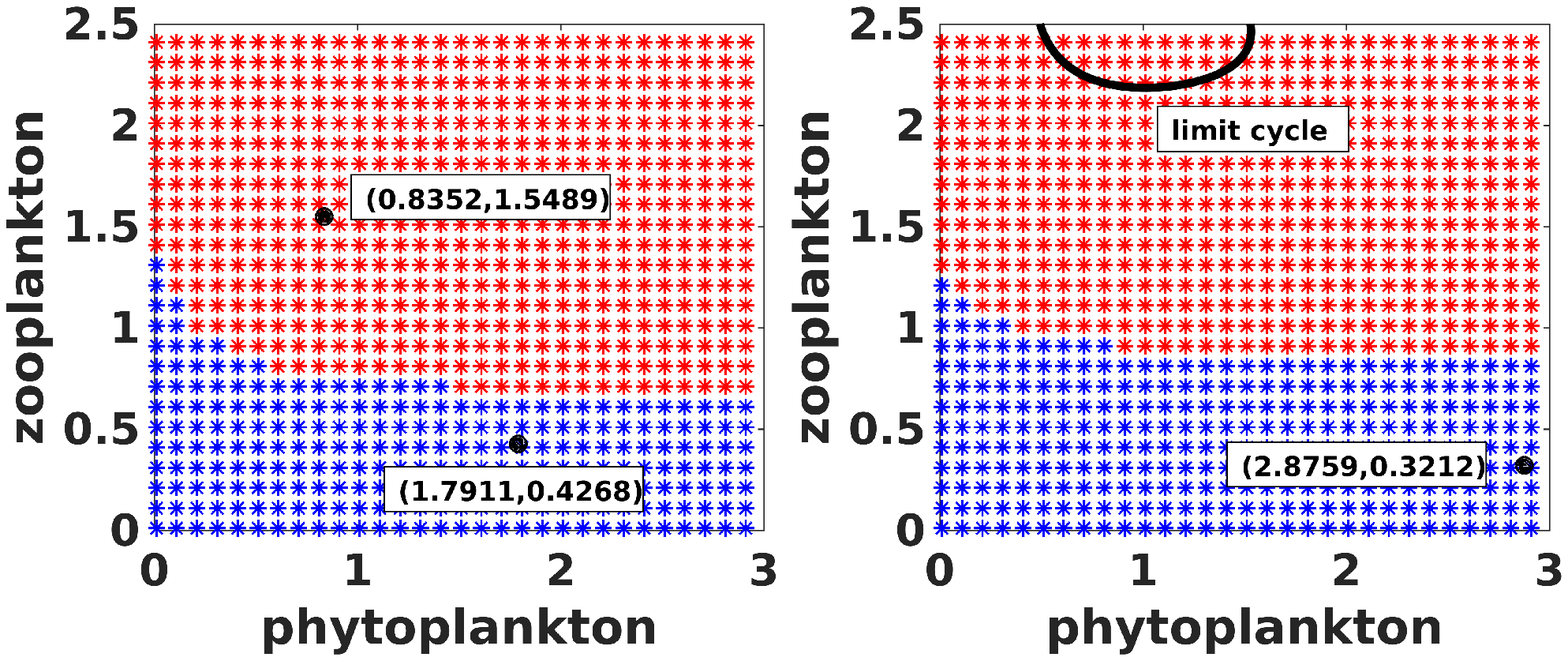}}
\end{center}
\caption{Basin of attraction for the bistable scenario for different carrying capacities. Left panel: $K=2$, $E=14.3$ $\mu g L^{-1}$; Right panel: $K=3$, $E=16$ $\mu g L^{-1}$. The red and the blue points denote initial conditions for which the system converges to the zooplankton dominated and phytoplankton dominated equilibrium respectively}
\label{boa}
\end{figure}

%
%

\bibliographystyle{spbasic}      
\bibliography{ref_cu}   


\end{document}